\documentstyle[eqsecnum,aps]{revtex}

\begin{document}
\draft \preprint{HEP/123-qed}
\title{Path Integral of the Holstein Model with a $\phi^4$ on site
potential}
\author{ Marco Zoli }
\address{Istituto Nazionale Fisica della Materia - Dipartimento di Fisica
\\ Universit\'a di Camerino, 62032, Italy. - marco.zoli@unicam.it}

\date{\today}

\maketitle
\begin{abstract}
We derive the path integral of the semiclassical, one dimensional
anharmonic Holstein model assuming that the electron motion takes
place in a bath of non linear oscillators with quartic on site
hard (and soft) potentials. The interplay between {\it e-ph}
coupling and anharmonic force constant is analysed both in the
adiabatic and antiadiabatic regime. In the latter we find much
larger anharmonic features on the thermodynamic properties of low
energy oscillators. Soft on site potentials generate attractive
centres at large amplitude oscillator paths and contribute to the
anomalous shape of the {\it heat capacity over temperature} ratio
in the intermediate to low $T$ range. This anharmonic lattice
effect is superimposed to the purely electronic contribution
associated to a temperature dependent hopping with variable range
inducing local disorder in the system.
\end{abstract}


\narrowtext
\section*{I. Introduction}

There is at present a large interest on the effects of a strong
electron-phonon coupling in a number of systems ranging from dimer
molecular junctions \cite{kaat} to carbon nanotubes \cite{mah},
from organic molecular crystals \cite{han}, to DNA \cite{alexan}
and cuprate superconductors \cite{sch,lan}. Several theoretical
studies have focussed on the interplay between e-ph coupling and
non linearities in the framework of the Holstein model
\cite{chris,voulga} investigating the phase diagram both in the
adiabatic \cite{zieg} and the antiadiabatic regime. Mainly in the
latter anharmonic effects are believed to be large \cite{pietro}
thus offering a picture to explain the high $T_c$ of binary alloys
such as superconducting $MgB_2$ with a small Fermi energy and
sizeable {\it e-ph} coupling \cite{cava}.

The path integral formalism provides a powerful method to study
quantum systems in which a particle is non linearly coupled to the
environment \cite{fehi,kleinert,eck}. A previous path integral
analysis \cite{io5} has pointed out how the phonon dispersion,
that has to be taken into account in the computation of the ground
state properties of the Holstein Hamiltonian
\cite{holst1,raedt2,io98}, induces non local {\it e-ph}
correlations which renormalize downwards the effective coupling
and ultimately broaden the size of the polaronic quasiparticle.
This explains why the polaron mass in a dispersive Holstein model
\cite{atin} is lighter than in a dispersionless model
\cite{alekor}. Also the thermodynamics of the Holstein Hamiltonian
can be computed within a dispersive model which accounts for the
lattice structure \cite{io5}.

The Holstein diatomic molecular model was originally cast
\cite{holst} in the form of a discrete non linear Schr\"odinger
equation for electrons whose probability amplitude at a molecular
site depends on the interatomic vibration coordinates. The non
linearities are tuned by the {\it e-ph} coupling \cite{kenkre}
whose strength drives the crossover between a large and a small
polaron state according to the degree of adiabaticity and the
dimensionality of the system \cite{trug}.

In the Holstein model, the phonon thermodynamics is not affected
by {\it e-ph} induced anharmonicities \cite{io4}. This follows
from the fact that the Holstein perturbing source current is local
in time and it does not depend on the electron path coordinate. As
a consequence, in the total partition function, electron and
lattice degrees of freedom are disentangled and the latter can be
integrated out analytically as far as a harmonic lattice model is
assumed. However non linearities may arise in Holstein like
systems also by virtue of on site potentials dependent on the
lattice structure and, in principle, independent of the {\it e-ph}
coupling. We are thus led to investigate the thermodynamics of the
anharmonic Holstein model with a quartic on site potential which
may be repulsive or partly attractive according to the sign of the
force constant and the amplitude of the lattice displacement
paths. The path integral approach permits to monitor the physical
properties for any value of the coupling strengths \cite{feynman}.
The presence of a $\phi^4$ potential may in turn affect also the
{\it e-ph} interactions and sinergically interfere on the
equilibrium properties of the system. This is the focus of the
present paper. Section II presents the Hamiltonian model while the
path integral method is briefly described in Section III. The
derivation of the total partition function of the system is
presented in Section IV and Section V contains the discussion of
the physical results. The conclusions are drawn in Section VI.

\section*{II. The Anharmonic Holstein Model}

We consider the one dimensional anharmonic Holstein Hamiltonian
consisting of: i) one electron hopping term, ii) an interaction
which couples the electronic density ($f_{\bf l}^{\dag}f_{\bf l}$)
to the lattice displacement $u_l$ at the {l}-site, iii) a bath of
$N$ identical dispersionless anharmonic oscillators with mass $M$
and frequency $\omega$ :

\begin{eqnarray}
& &H =\, H^e + H^{e-ph} + H^{ph} \nonumber
\\
& &H^e =\, - t \sum_{<{l, m}>} f_{l}^{\dag} f_{m} \nonumber
\\
& &H^{e-ph}=\,  {\bar g} \sum_{l=1}^Nu_l f^{\dag}_l f_{l} \,
\nonumber
\\
& &H^{ph}=\, {M \over 2} \sum_{l=1}^N \bigl(\dot{u}_l^2 + \omega^2
u_l^2 \bigr) + {{\delta} \over 4} \sum_{l=1}^N u_l^4  \, \nonumber
\\ & &{\bar g}=\,g \sqrt{2M \omega}
\end{eqnarray}

the sum $<,>$ is over $z$ nearest neighbors and $t$ is the tight
binding overlap integral.  $g$ is the {\it e-ph} coupling in units
of $\hbar \omega$. Choosing the atomic mass $M$ of order $10^4$
times the electron mass, we get ${\bar g} \simeq \, 1.1456 \times
g \sqrt{2 \hbar \omega}{}\, \bigl[ meV \AA^{-1} \bigr]$ where
$\hbar \omega$ is given in $meV$. $\delta$, in units $meV
\AA^{-4}$, controls the strength of the non linearities and
determines whether the on site potential $V(u_l)=\, {M}\omega^2
u_l^2/2 + {{\delta}}u_l^4/4$ is hard ($\delta > 0$) or soft
($\delta < 0$). In the latter case, $V(u_l)$ attains the maximum
at $u_l^2=\,M\omega^2/|\delta|$ and the inflection point occurs at
$u_l^2=\,M\omega^2/3|\delta|$. The condition $|\delta|u_l^2 > 2
M\omega^2$ yields an attractive on site potential. Then, the range
of the atomic path amplitudes generating attractive scattering
centres depends on the value of the anharmonic force constant. For
$|\delta| > 2 M\omega^2$ the potential becomes attractive for a
portion of large amplitude atomic paths while small amplitude
paths weigh the repulsive range.

In the following computation of the electron path integral coupled
to the anharmonic oscillator, after setting the potential
parameters, we select at any temperature the class of atomic paths
which mainly contribute to the euclidean action. As the
distribution of the path amplitudes has a cutoff on the scale of
the lattice constant, say i.e. $u_l^2 < 1 \AA^2$, the on site
potentials are always bound from below also in the attractive
cases.

\section*{III. The Path Integral Method}

The Holstein Hamiltonian in (1) can be mapped onto the time scale
according to space-time mapping techniques extensively described
in previous works \cite{io5,hamann,io3} and hereafter outlined.

Defining ${x}(\tau)$ and ${y}(\tau')$ as the electron coordinates
at the ${l}$ and ${m}$ lattice sites respectively, $H^e$ in (1)
transforms into

\begin{equation}
H^e(\tau,\tau')=\, -{t} \bigl( f^{\dag}({x}(\tau)) f({y}(\tau')) +
f^{\dag}({y}(\tau')) f({x}(\tau)) \bigr)
\end{equation}

where $\tau$ and $\tau'$ are continuous variables $\bigl( \in [0,
\beta]\bigr)$, with $\beta$ being the inverse temperature. After
setting $\tau'=\,0$, ${y}(0) \equiv 0$ and taking the thermal
averages for the electron operators over the ground state of the
Hamiltonian one gets the average electron hopping energy per
lattice site:

\begin{eqnarray}
& &h^e(x(\tau)) \equiv {{<H^e(\tau)>} \over N}= \, - {t}\Bigl(G[-{
x}(\tau), -\tau ] + G[{x}(\tau), \tau ]\Bigr)\, \nonumber \\
\end{eqnarray}

where, $G[{x}(\tau), \tau ]$ is the electron propagator at finite
temperature.

By treating the lattice displacements in (1) as $\tau -$ dependent
classical variables, $u_l \rightarrow u(\tau)$, we obtain from
$H^{e-ph}$ in (1) the averaged {\it e-ph} energy per lattice site
which is identified as the perturbing source current $j(\tau)$ in
the path integral method:

\begin{eqnarray}
j(\tau) \equiv {{<H^{e-ph}(\tau)>} \over N}= \,{\bar g} u(\tau)
\end{eqnarray}

As the Hamiltonian model assumes a set of identical oscillators we
study the path integral for the electron coupled to a single
anharmonic oscillator of the bath. The path integral reads:

\begin{eqnarray}
& &<{x}(\beta)|{x}(0)> =\, \int D{x}(\tau) exp\Biggl[-
\int_0^{\beta}d\tau E(x(\tau))\Biggr] \nonumber
\\ &\times& \int Du(\tau) exp\Biggl[- \int_0^{\beta}
d\tau O(u(\tau)) \Biggr] \,\nonumber
\\
& &E(x(\tau))=\,{{m_e} \over 2} \dot{{x}}^2(\tau) + h^e(x(\tau))
\,\nonumber
\\
& &O(u(\tau))=\, {M \over 2} \bigl( \dot{u}^2(\tau) + \omega^2
u^2(\tau) \bigr) + {{\delta} \over 4} u^4(\tau) +
j(\tau)\,\nonumber
\\
\end{eqnarray}

where the kinetic term ($m_e$ is the electron mass) is normalized
by the functional measure of integration over the electron paths.

Since the electron hopping does not induce a shift of the
oscillator coordinate the Holstein {\it e-ph} interactions are
local in time and, in the semiclassical treatment, the source
current $j(\tau)$ is independent of the electron path. As a
consequence oscillator and electron coordinates appear
disentangled in (5) while the coupling occurs through the
parameter $\bar{g}$.

\section*{IV. The Partition Function}

The quantum statistical partition function $Z_T$ is derived by
integrating (5) after imposing periodicity conditions, $\beta$ is
the period, both on the electron and oscillator paths:

\begin{eqnarray}
Z_T &=&\,\int dx <{x}(\beta)|{x}(0)> =\, Z_{el} \times Z_{osc}
\nonumber
\\
Z_{el} &=&\, \oint Dx(\tau) exp \Bigl[- \int_0^{\beta}d\tau
E(x(\tau)) \Bigr] \nonumber
\\
Z_{osc} &=&\, \oint Du(\tau) exp\Bigl[- \int_0^{\beta} d\tau
O(u(\tau)) \Bigr]
\,\nonumber \\
\end{eqnarray}

where  $\oint Dx(\tau)$ and $\oint Du(\tau)$ are the functional
measures of integration.

The electronic contribution $Z_{el}$ is computed by expanding the
paths in Fourier components

\begin{eqnarray}
x(\tau)&=&\,x_o + \sum_{m=1}^{M_F}\bigl( r_m \cos(\nu_m \tau) +
s_m \sin(\nu_m \tau) \bigr)\, \nonumber
\\
&&r_m =\,2\Re x_m \, \nonumber
\\
&&s_m =\, -2\Im x_m \, \nonumber
\\
&&\nu_m =\,2m\pi/\beta \, \nonumber
\\
\end{eqnarray}

and taking the following measure of integration

\begin{eqnarray}
&& \oint Dx(\tau) \equiv {{\sqrt{2}} \over {(2
\lambda_{m_e})^{(2M_F+1)}}} \int_{-\infty}^{\infty}{dx_o} \,
\nonumber
\\
&\times&
\prod_{m=1}^{M_F}  (2\pi m)^2 \int_{-\infty}^{\infty} dr_m
\int_{-\infty}^{\infty} ds_m \, \nonumber
\\
& &\lambda_{m_e}=\,\sqrt{\pi \hbar^2 \beta/m_e}\, \nonumber
\\
\end{eqnarray}

The cutoffs over the Fourier coefficient integrations have to
ensure proper normalization of the kinetic term in absence of
hopping processes. Thus $Z_{el}$ transforms into

\begin{eqnarray}
Z_{el}&\simeq& \,{{\sqrt{2}} \over {(2 \lambda_{m_e})^{(2M_F+1)}}}
\int_{-\Lambda/2}^{\Lambda/2}{d{x}_o} \prod_{m=1}^{M_F}
\int_{-\Lambda}^{\Lambda} d{r}_m \int_{-\Lambda}^{\Lambda} d{s}_m
\, \nonumber
\\ &\times& \exp\biggl(- {{\pi^{3}} \over \lambda^2_{m_e}} \sum_{m=1}^{M_F}
m^2({r}^2_m + {s}^2_m) - \int_0^{\beta}d\tau h^e(x(\tau)) \biggr)
\, \nonumber
\\
\end{eqnarray}

with $\Lambda \propto \lambda_{m_e}$ \cite{io3} indicating that
large amplitude electron paths have to be selected at low
temperatures where the quantum effects are larger. Two Fourier
components, $M_F=\,2$, suffice to attain stable results as
$h^e(x(\tau))$ depends smoothly on the electron path. The hopping
term accounts for the deviation from the Gaussian behavior.
Numerical analysis shows that, for any choice of path parameters,
$h^e(x(\tau))$ decreases by decreasing the temperature but its
overall contribution to the electron action is substantial also at
low $T$.

Let's focus now on the anharmonic oscillator term $Z_{osc}$. The
oscillator path is expanded in $N_F$ Fourier components

\begin{eqnarray}
u(\tau)=\,u_o + \sum_{n=1}^{N_F}\bigl( a_n \cos(\omega_n \tau) +
b_n \sin(\omega_n \tau) \bigr)\, \nonumber
\\
\end{eqnarray}

with Matsubara frequencies $\omega_n=\,2n\pi/\beta$ and
coefficients $a_n\equiv \Re u_n$, $b_n\equiv -\Im u_n$ satisfying
the conditions $a_n =\,a_{-n}$ and  $b_n =\,-b_{-n}$. The latter
are consistent with the choice of real paths and simplify the
following $\tau$ integration of the on site potential.

Note that the periodicity property, $u(\tau) =\,u(\tau + \beta)$,
would be fulfilled also taking the very $a_n$ coefficients in (10)
\cite{fey}. However such a choice would not permit to fit with
accuracy the harmonic oscillator partition function which is known
exactly: $Z_h =\,\bigl[2 \sinh(\beta\omega/2)\bigr]^{-1}$. Infact,
in the path integral method, the harmonic partition function
$Z^{PI}_h$ reads

\begin{eqnarray}
Z^{PI}_h =\, \oint Du(\tau) exp\Biggl[- \int_0^{\beta} d\tau
\Bigl[ {M \over 2} \bigl( \dot{u}^2(\tau) + \omega^2 u^2(\tau)
\bigr) \Bigr] \Biggr]\, \nonumber
\\
\end{eqnarray}

and taking the functional measure

\begin{eqnarray}
\oint Du(\tau) \equiv & & {{\sqrt{2}} \over {(2
\lambda_M)^{(2N_F+1)}}} \int_{-\infty}^{\infty}{du_o} \,
\nonumber
\\
&\times& \prod_{n=1}^{N_F}  (2\pi n)^2 \int_{-\infty}^{\infty}
da_n \int_{-\infty}^{\infty} db_n \, \nonumber
\\
\end{eqnarray}

with $\lambda_M=\,\sqrt{\pi \hbar^2 \beta/M}$, one gets from
(10)-(12):

\begin{eqnarray}
Z^{PI}_h=\, {1 \over {\beta \omega}} \prod_{n=1}^{N_F}
{{(2n\pi)^2} \over {{(2n\pi)^2 + (\beta \omega)^2}}} \, \nonumber
\\
\end{eqnarray}

Instead, dropping the $b_n$ terms in (10) and (12), one would get
the square root of the product series in (13) which does not yield
a reliable fit of $Z_h$ even for large $N_F$. Note that at high
$T$, the condition $2n\pi \gg \beta \omega$ is fulfilled for small
integers $n$, hence the main contribution to $Z^{PI}_h$ is given
by the ${1 /{\beta \omega}}$ factor which stems from the
$\int{du_o}$ in (12). This is consistent with the expectation that
high $T$ paths are well approximated by their $\beta$-averaged
value $u_o$ whereas fluctuation effects become increasingly
relevant towards the low $T$ regime in which $N_F$ rapidly grows.
$N_F$ clearly varies also with the oscillator energy while the
shape of $N_F(\omega,T)$ may differ according to the harmonic
function ($Z_h$, harmonic free energy or specific heat) one
chooses to fit.

The anharmonic partition function $Z_{osc}$ in (6) can be worked
out analytically using (10),(12) and performing the time
integration of the oscillator functional $O(u(\tau))$. This
permits to get an insight into the role of the non linear terms.
The lenghty calculation yields:

\begin{eqnarray}
&&Z_{osc}=\, {{\sqrt{2}} \over {(2 \lambda_M)^{(2N_F+1)}}}
\int_{-\infty}^{\infty}{du_o} \exp\bigl( -\beta \bar g u_o -
\kappa u_o^2 -\beta\delta u_o^4/4 \bigr)\, \nonumber \\ && \times
\prod_{n=1}^{N_F} (2\pi n)^2 \int_{-\infty}^{\infty} da_n
\exp\Biggl[-\sum_{n=1}^{N_F} \Bigl[ (\gamma_n + 3\beta\delta
u_o^2/4)a_n^2  \nonumber
\\
&&+ {{\beta\delta u_o} \over 2} \sum_{m=1}^{N_F}c({n,m}) +
{{\beta\delta} \over {16}} \sum_{m,p=1}^{N_F}d({n,m,p}) \Bigr]
\Biggr]\, \nonumber \\ && \times \int_{-\infty}^{\infty} db_n
\exp\Biggl[-\sum_{n=1}^{N_F} \Bigl[ (\gamma_n + 3\beta\delta
u_o^2/4)b_n^2  \nonumber
\\
&&+ {{3\beta\delta u_o} \over 2} \sum_{m=1}^{N_F}e({n,m}) +
{{\beta\delta} \over {16}} \sum_{m,p=1}^{N_F}\bigl({{6}}f({n,m,p})
- g({n,m,p}) \bigr) \Bigr] \Biggr]\, \nonumber \\ \nonumber
\\
&&\kappa=\,\pi
(\beta\omega)^2/2\lambda_M^2 \, \nonumber
\\
&& \gamma_n=\, \pi \bigl((2\pi n)^2 + (\beta\omega)^2 \bigr)/4
\lambda_M^2 \, \nonumber
\\
&& c({n,m})=\,a_n a_m (a_{n+m}+a_{n-m}) \nonumber
\\
&& d({n,m,p})=\,a_n a_m a_p
(a_{n+m+p}+a_{n-m+p}+a_{p-n-m}+a_{p-n+m}) \nonumber
\\
&& e({n,m})=\,a_n b_m (b_{n+m}- b_{n-m}) \nonumber
\\
&& f({n,m,p})=\,a_n a_m b_p
(b_{n+m+p}+b_{n-m+p}+b_{p-n-m}+b_{p-n+m}) \nonumber
\\
&& g({n,m,p})=\,b_n b_m b_p
(b_{n+m+p}-b_{n-m+p}+b_{p-n-m}-b_{p-n+m}) \nonumber
\\
\end{eqnarray}

In $c({n,m})$ the coefficients $a_{j}$ $(j=\, n+m, n-m)$ vanish
if: $j \leq 0$ or $j > N_F$. In $d({n,m,p})$, $a_k \neq 0 \,\,
(k=\,n+m+p, \,\, n-m+p, \,\, p-n-m, \,\, p-n+m)
\Longleftrightarrow 1 \leq k \leq N_F$. Analogous conditions hold
for the coefficients $b_{j\, (k)}$ in the $e({n,m})$, $f({n,m,p})$
and $g({n,m,p})$ functions.

Note that the effective {\it e-ph} coupling $\bar{g}$ is
associated only to the $\tau$-independent component $u_o$, that is
to the $\beta$-averaged displacement path \cite{io5}. This follows
from the fact that the perturbing source current is non retarded
in the Holstein model as a consequence of the local nature of the
e-ph interaction.

The quartic potential induces a strong mixing of the Fourier
components of the path that highly complicates the numerical
problem. Thus the value of $N_F$ appears to be crucial in the
computation. We determine $N_F(\omega,T)$ by fitting (with an
accuracy of $2 \cdot 10^{-2}$) the exact harmonic free energy
($F_h=\,-\ln(Z_h)/\beta$), through the path integral harmonic free
energy $F_h^{PI}$ obtained from (13). As an example, for the
oscillator with $\omega=\,20meV$, $N_F(T=\,10K)=\,59$ and
$N_F(T=\,200K)=\,8$.

Inspection of (14) offers the key to perform reliable path
integrations according to the sign of the $\phi^4$ potential. At
high temperature, a large contribution to $Z_{osc}$ is expected to
come from the paths having $u_o$ which maximizes $\exp\bigl(\kappa
f(u_o)\bigr)$ with $f(u_o)=\, -(a u_o + u_o^2 + b u_o^4), \, a
\equiv \,\beta \bar g/\kappa; \,\, b \equiv\,\beta\delta/4\kappa$.
In general, we find that for a hard (soft) potential, the
$du_o$-integration has to be carried out along the $u_o < 0 \,(u_o
> 0)$ axis, with cutoff $|u_o|_{max} \sim 0.6/\sqrt{\kappa}$. This
permits to include the set of paths which mainly contribute to the
euclidean oscillator action. On the Fourier coefficients
integrals, $\int da_n \,\int db_n$, we set the cutoffs
$|a_n|_{max}, \, |b_n|_{max} \sim 0.6/\sqrt{\gamma_n}$ both for
hard and soft potentials thus achieving numerical convergence and
correct computation of the Gaussian integrals once the non
linearities are switched off. It turns out, see the definitions in
(14), that the cutoffs on the oscillator path integration are
increasing functions of temperature ($\propto \sqrt{T}$)
consistently with the physical expectations of large amplitude
displacements at high $T$. As the path displacements encounter an
upper limit due to the cutoffs, $u(\tau) \leq |u_o|_{max} +
2\sum_{n=1}^{N_F}|a_n|_{max}$, the distribution of on site
potentials has a lower limit also in the case of soft and
attractive non linearities. This avoids numerical divergences and
makes the problem physically meaningful.

\section*{V. Results}

We test the relevance of the non linearities on the equilibrium
thermodynamics of the system and present the calculation for the
heat capacity in the intermediate to low temperature range.

Figures 1 show the behavior of a low energy ($\omega=\,20meV$)
oscillator without ($g=\,0$) and with ($g=\,2,4$) coupling to the
electronic subsystem in the adiabatic regime: $t/ \omega =\,5$.
Figs.1(a),(b) assume an anharmonic potential with positive quartic
force constant $\delta = 10^3, 10^4 meV \AA^{-4}$ respectively.
Also the harmonic heat capacity is reported on for comparison. The
hard potential lifts the free energy over the harmonic values with
a more pronounced enhancement at increasing $T$ and for larger
$\delta$. The effect on the free energy second derivative is
however scarce and essentially consists, see Fig.1(a), in a slight
increase ( decrease) of the heat capacity at low (high)
temperatures. The reduction of the heat capacity (with respect to
the harmonic plot) at intermediate and high $T$ is more evident in
Fig.1(b) where the quartic force constant is larger. This result
is in accordance with diagrammatic perturbative treatments of the
anharmonic crystals which predict negative contributions to the
constant volume specific heat arising from positive force
constants in the quartic potential \cite{martin,io90}. The {\it
e-ph} coupling strength also tends to decrease the oscillator heat
capacity but the overall effect is small as the insets in the
figures show: infact the electronic term dominates the total heat
capacity $C$ and the oscillator contribution is not
distinguishable in the plots of the $C$ over temperature ratios
versus $T$. The low temperature upturn is due to the large
electron energy term in (9) and precisely ascribable to the
feature of the variable range (on the $\tau$ scale) of the
electron hopping, captured by the path integral formalism
\cite{io3}.

The cases of a soft on site potential are reported on in Fig.1(c)
with $\delta = -10^3 meV \AA^{-4}$ and Fig.1(d) with $\delta =
-10^4 meV \AA^{-4}$. The characteristic potential parameter is
$M\omega^2/|\delta| \simeq 0.5$ and $0.05 \AA^2$, respectively.
The potential $V(u(\tau))=\, {M}\omega^2 u(\tau)^2/2 +
{{\delta}}u(\tau)^4/4$ is attractive for those paths such that
$u(\tau)^2 > 2M\omega^2/|\delta|$.

At any $T$, we integrate over a distribution of time dependent
potentials. Thus, at a given lattice site, the electron may
experience an attractive or repulsive scattering centre according
to the size of the atomic path. As an example, at $T=\,200K$, we
get a maximum path $u_{max}$ such that $u_{max}^2 \sim 0.15
\AA^2$. This guarantees that $V(u(\tau))$ is generally repulsive
in Fig.1(c) and attractive for a broad class of paths in Fig.1(d).

The effects on the oscillator heat capacity are twofold: i) the
soft potential enhances the heat capacity mainly in the low $T$
range with respect to the hard potential and this feature is much
more pronounced in Fig.1(d); ii) the trend of the {\it e-ph}
coupling is opposite to that observed in Figs.1(a),(b): now by
increasing the $g$ values one gets higher heat capacities although
the size of this effect is small on the scale of the electronic
terms in the adiabatic regime as revealed by the $C/T$ plots in
the insets.

Let's come to the antiadiabatic regime ($t/ \omega =\,0.5$)
discussed in Figures 2, where a low harmonic energy ($\omega
=\,10meV$) is assumed to emphasize the size of the anharmonicity
together with a very narrow electron band. A hard on site
potential with $\delta = 10^3 meV \AA^{-4}$ is taken in Fig.2(a):
the shape of the oscillator heat capacity signals the effects of
the non linearities which flatten the curve at intermediate $T$
and enhance the heat capacity also at low $T$ with respect to the
corresponding case of Fig.1(a). As the Debye temperature is now
smaller (than in Fig.1(a)) the hard anharmonicity decreases the
constant volume heat capacity with respect to the harmonic plot
over a broader temperature range.

The {\it e-ph} coupling plays a minor role also in antiadiabatic
conditions. The anharmonic contribution is visible in the total
heat capacity as the inset makes evident although the dominant
electronic feature persists in the low $T$ upturn of $C/T$. In
Fig.2(b), we assume $g =\,2$ and consider two cases of soft
potential: the oscillator anharmonicity becomes relevant and such
to modify the shape of the anomalous upturn in the total heat
capacity. An enhancement of the $C/T$ values is observed at
intermediate and low $T$ and, in the case of the largest
$|\delta|$ generating a soft attractive potential, the oscillator
heat capacity $C_{osc}$ yields an upturn in $C_{osc}/T$
independently of the electronic term.

\section*{VI. Conclusions}

We have studied the path integral of the one dimensional non
linear Holstein model in which a set of dispersionless oscillators
provides the environment for the electron. The model is
semiclassical as the lattice displacements are treated classically
while the electron operators are thermally averaged over the
ground state Hamiltonian. The {\it e-ph} coupling of the model is
local and generates a perturbing current which linearly depends on
the oscillator path amplitude $u(\tau)$ where $\tau$ is the time
(or inverse temperature) of the Matsubara Green functions
formalism. The anharmonicity on the lattice site is modelled
through a $\phi^4$ potential that may result attractive for a set
of displacement paths in the case of a negative quartic force
constant (soft potential). We have derived the path integral of
the interacting system and computed the total partition function
selecting, as a function of the temperature ($T \leq 200K$), both
the electron and oscillator paths which yield the largest
contribution to the action. While quantum electron paths have
increasing amplitudes at decreasing temperatures, the atomic
displacements are growing functions of $T$. This relevant physical
feature is accounted for in our model as the cutoffs on the
electron path integration are proportional to the electron thermal
wavelength whereas,  on the atomic path integration, we find
cutoffs proportional to $\sqrt{T}$.

The oscillator partition function includes the effect of the
coupling to the electron subsystem but the on site anharmonicities
play a major role mainly when the potential is soft and the
harmonic energy is low. Among the thermodynamic properties we have
chosen to present the heat capacity $C$ in view of the upturn in
the $C/T$ behavior due to the low $T$ electron hopping tuned by
the value of the overlap integral in the Hamiltonian model. The
computation is highly time consuming especially because of the
strong mixing of the path Fourier components generated by the non
linear potential.

In general we find that: i) in the case of a hard quartic
potential, switching on the {\it e-ph} coupling leads to lower the
free energy and its second derivative, ii) when the quartic
potential is soft, {\it e-ph} coupling and anharmonicity act
sinergically enhancing the thermodynamic functions.

These results can be understood on general physical grounds.
Infact, a hard quartic potential shifts the characteristic phonon
frequency upwards thus hardening the spectrum and broadening the
size of the quasiparticle. If the {\it e-ph} coupling is enhanced
(independently of the on site anharmonicity) the magnitude of the
source which causes the lattice distortion becomes larger. This
further hardens the vibrations and leads to decrease the
anharmonic heat capacity over the whole temperature range.
Instead, when the quartic potential is soft the phonon frequency
is lowered and the oscillator potential well is more flexible. In
this case, larger {\it e-ph} coupling strengths favour the
self-trapping of quasiparticles with heavier effective masses.
This is physically equivalent to soften the phonon spectrum and
enhance the heat capacity.

The electron contribution to the heat capacity is dominant in the
adiabatic regime whereas antiadiabatic systems are expected to
present significant anharmonic corrections. Infact, in the
antiadiabatic regime the quasiparticle is a small size object on
the lattice scale and the electron energy associated with the
overlap integral is small. Thus, this regime proposes a physical
picture in which the electron hardly hops from site to site and
its effective mass becomes heavier on the scale of the atomic
mass. But a potential well generated by "lighter oscillators" is
more sensitive to on site anharmonic effects.

In particular, soft potentials increase the heat capacities over
the harmonic values and reinforce the upturn in the $C/T$ versus
$T$ plots when the on site anharmonicity is such to produce
attractive potentials for a set of lattice displacement paths.
Since the path amplitudes are larger at high $T$, soft attractive
potentials induce rapidly increasing phonon heat capacities at
growing $T$ as shown in Fig.2(b).

Thus our path integral investigation and the thermodynamical
results point to a complex role of the lattice anharmonicities in
the one dimensional Holstein model and suggest that on site
potentials may be experienced as attractive or repulsive according
to the temperature and the amplitude of the atomic path. Such
potentials may provide scattering centres generating local
disorder whose effect on the system thermodynamics is superimposed
to the disorder induced by the hopping of electrons with variable
range.

\begin{figure} \vspace*{18truecm}
\caption{(Color online) Anharmonic Oscillator Heat Capacities
versus temperature for three values of {\it e-ph} coupling $g$ and
oscillator energy $\omega =\,20meV$. (a) Hard potential with force
constant $\delta = 10^3 meV \AA^{-4}$; (b) $\delta = 10^4 meV
\AA^{-4}$; (c) soft potential with $\delta = -10^3 meV \AA^{-4}$;
(d) $\delta = -10^4 meV \AA^{-4}$. The harmonic heat capacity is
plotted in (a) and (b) for comparison. The insets show the {\it
Total (electronic plus anharmonic oscillator) Heat Capacity over
temperature} ratios in the adiabatic regime $t/ \omega =\,5$.}
\end{figure}

\begin{figure} \vspace*{12truecm}
\caption{(Color online) Anharmonic Oscillator Heat Capacities
versus temperature in the antiadiabatic regime $t/ \omega =\,0.5$
and oscillator energy $\omega =\,10meV$. The harmonic heat
capacity is also plotted. (a) Hard potential force constant
$\delta = 10^3 meV \AA^{-4}$ with three values of {\it e-ph}
coupling $g$. (b) Two soft potential force constants at fixed {\it
e-ph} coupling. The insets show the {\it Total (electronic plus
anharmonic oscillator) Heat Capacity over temperature} ratios. The
electronic contribution is plotted separately for comparison. }
\end{figure}

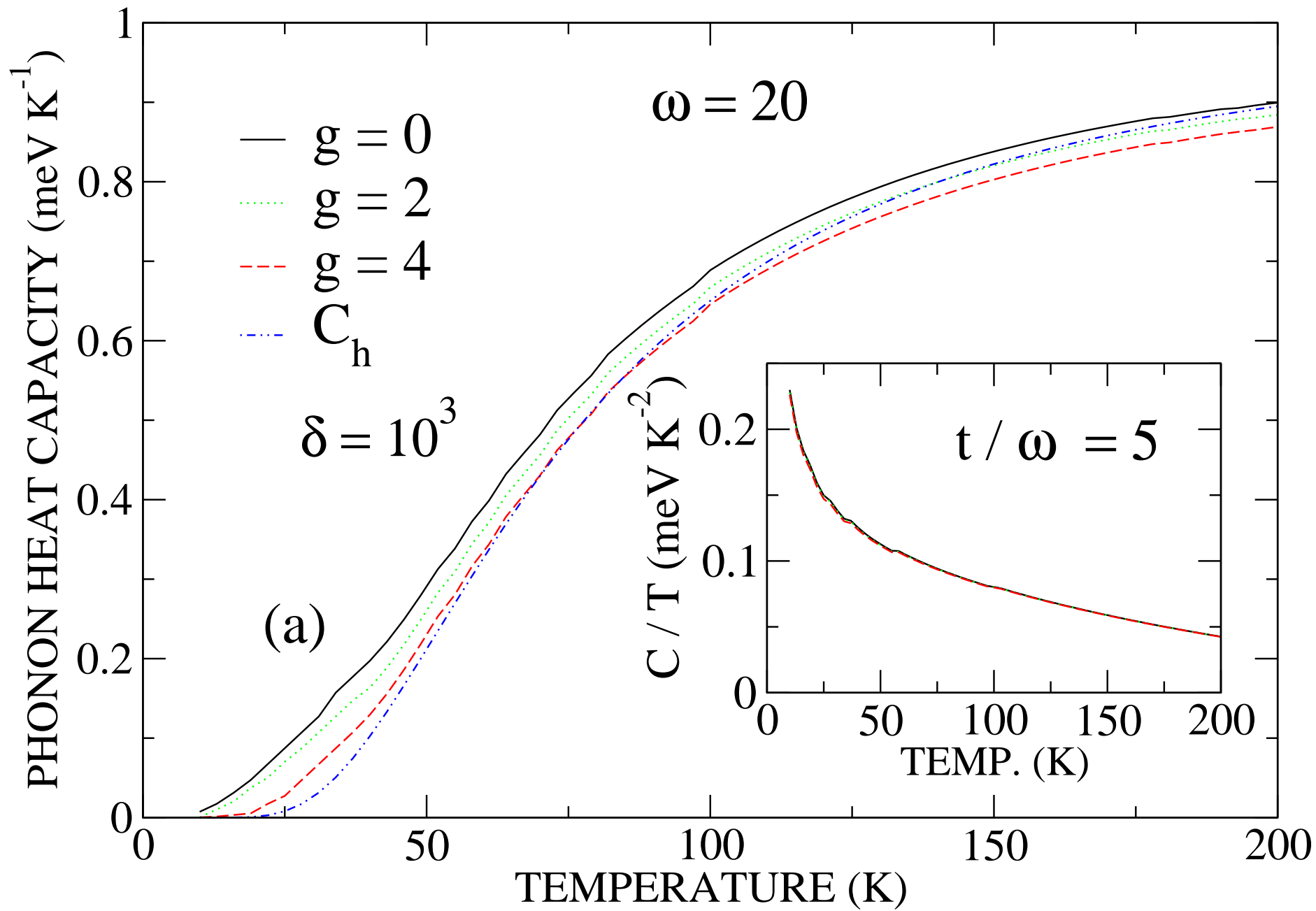
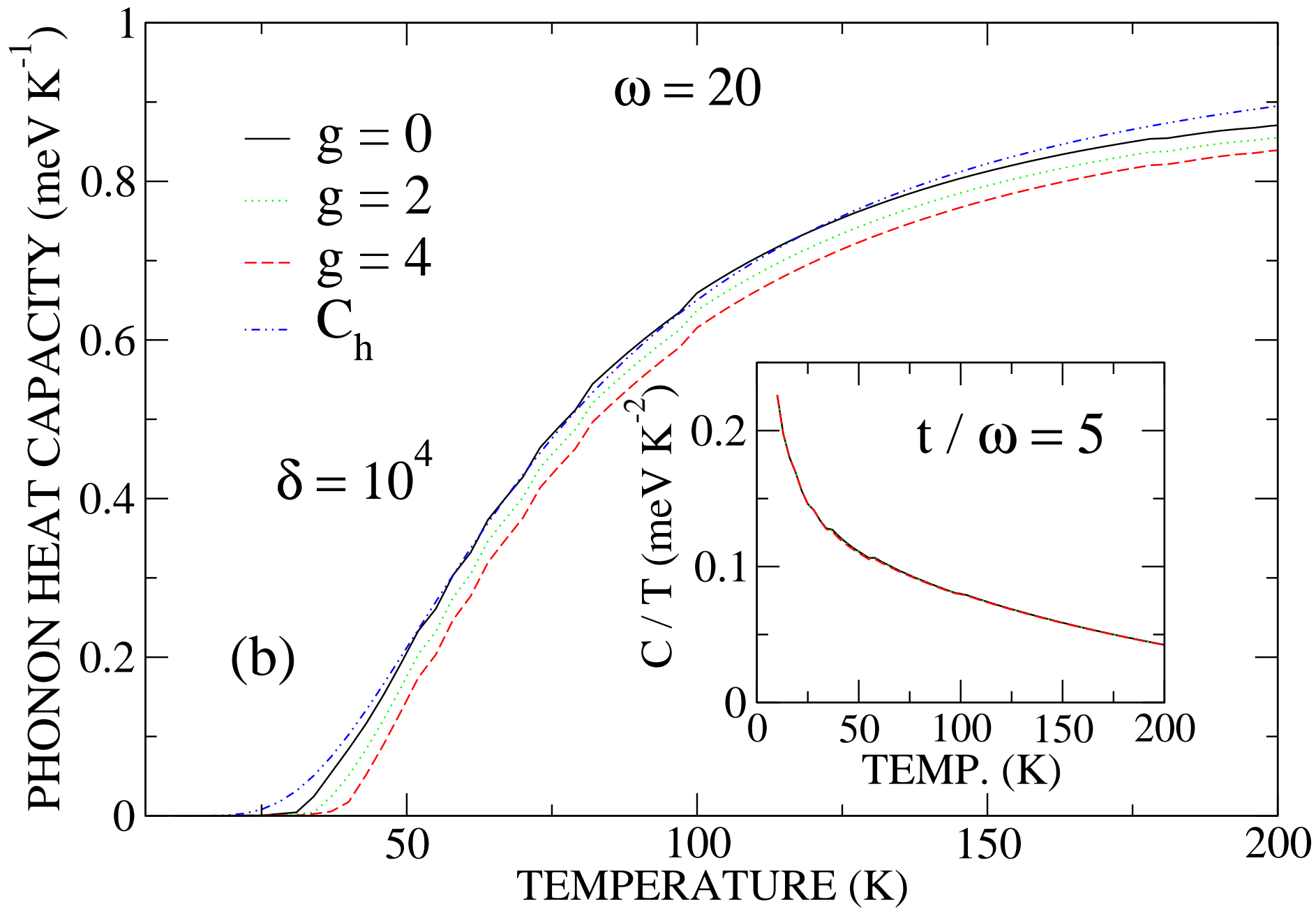
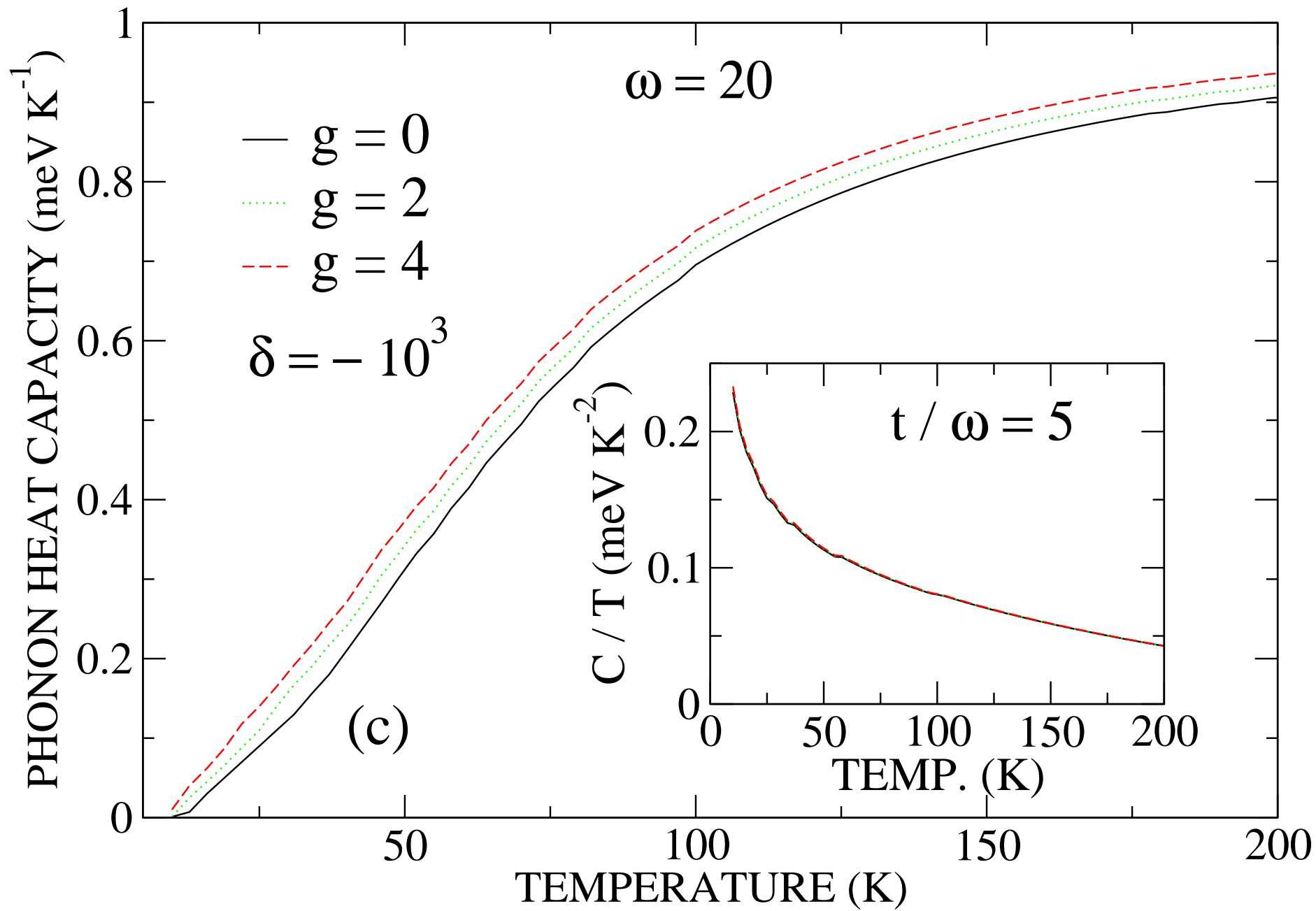
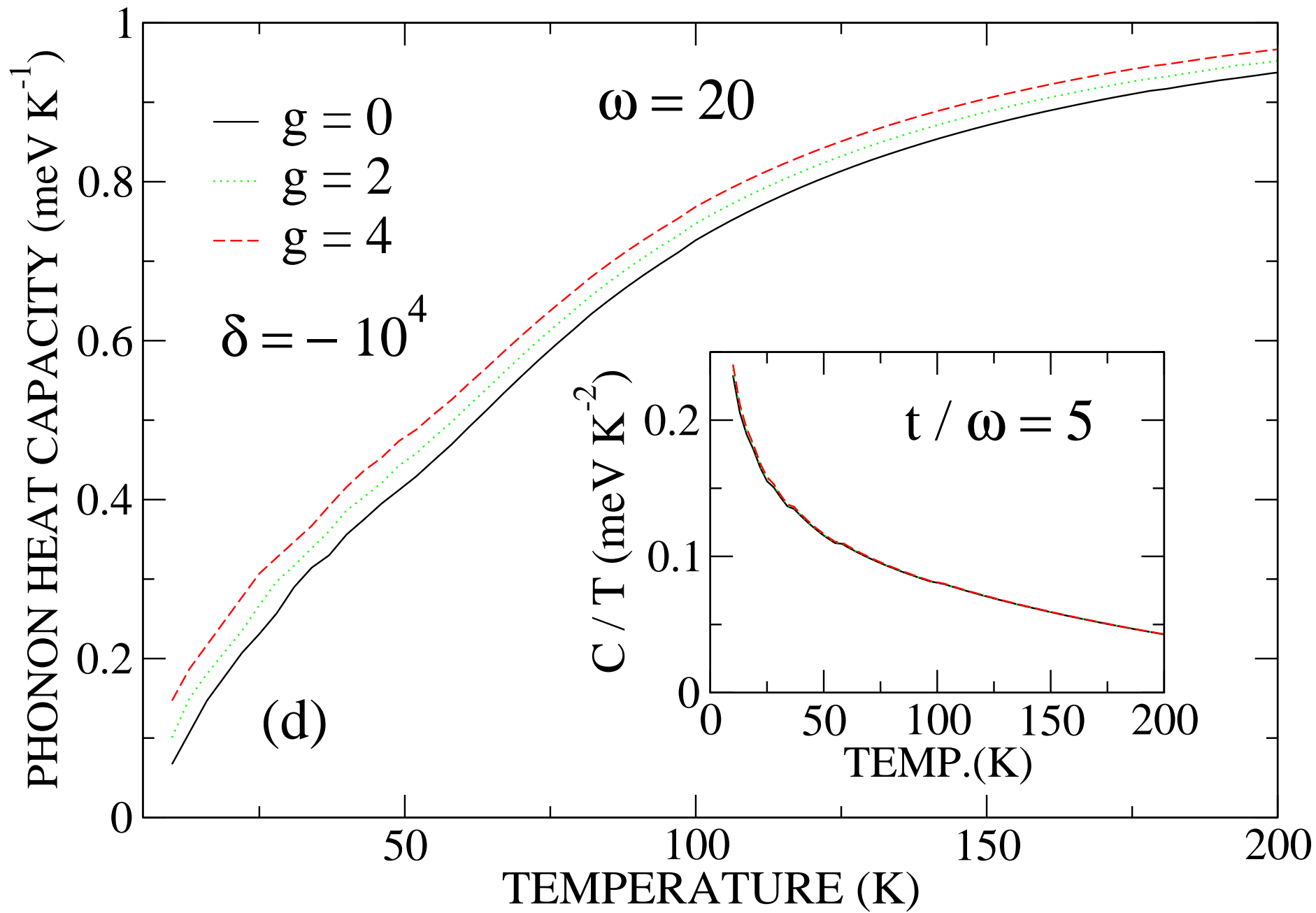
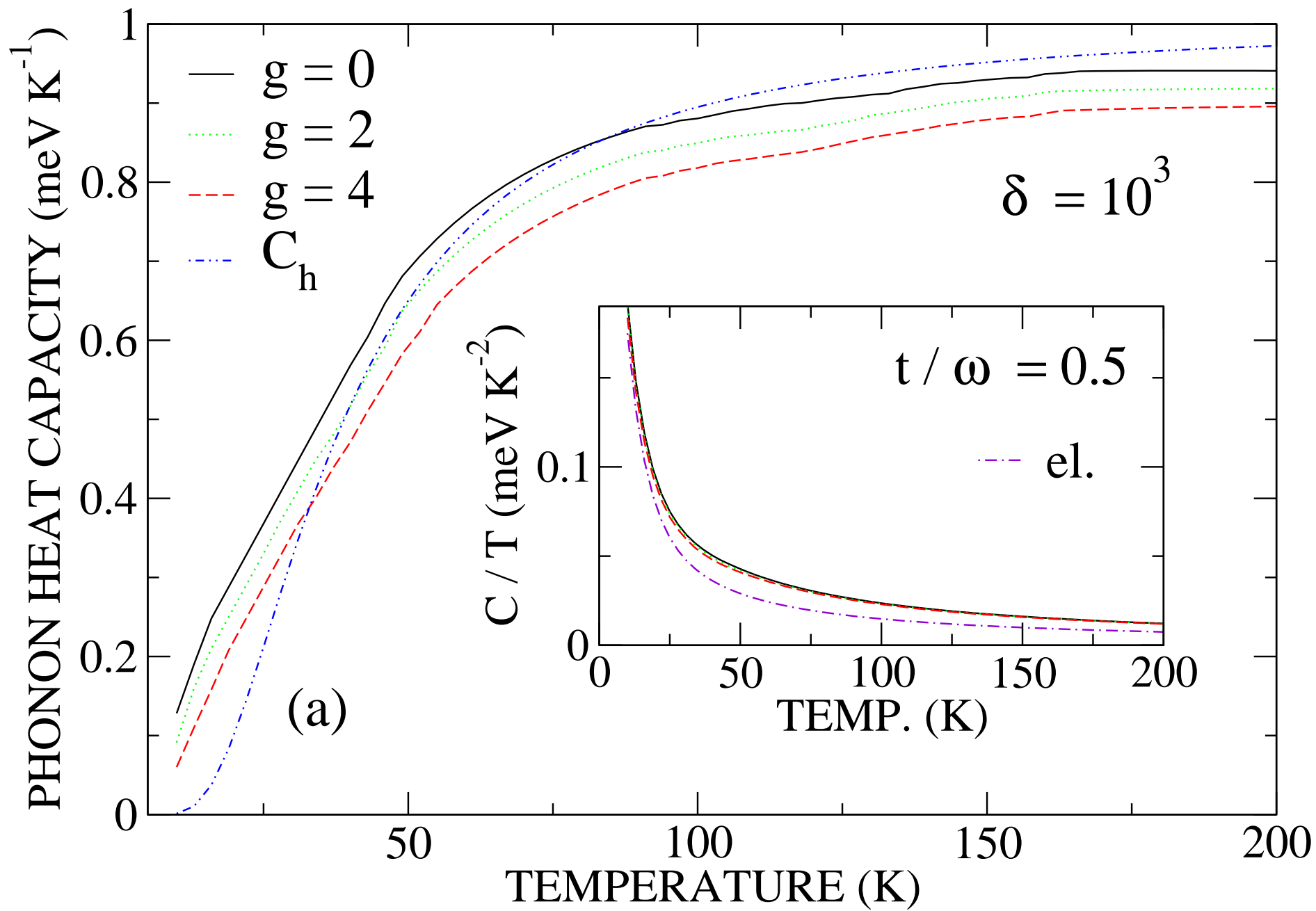
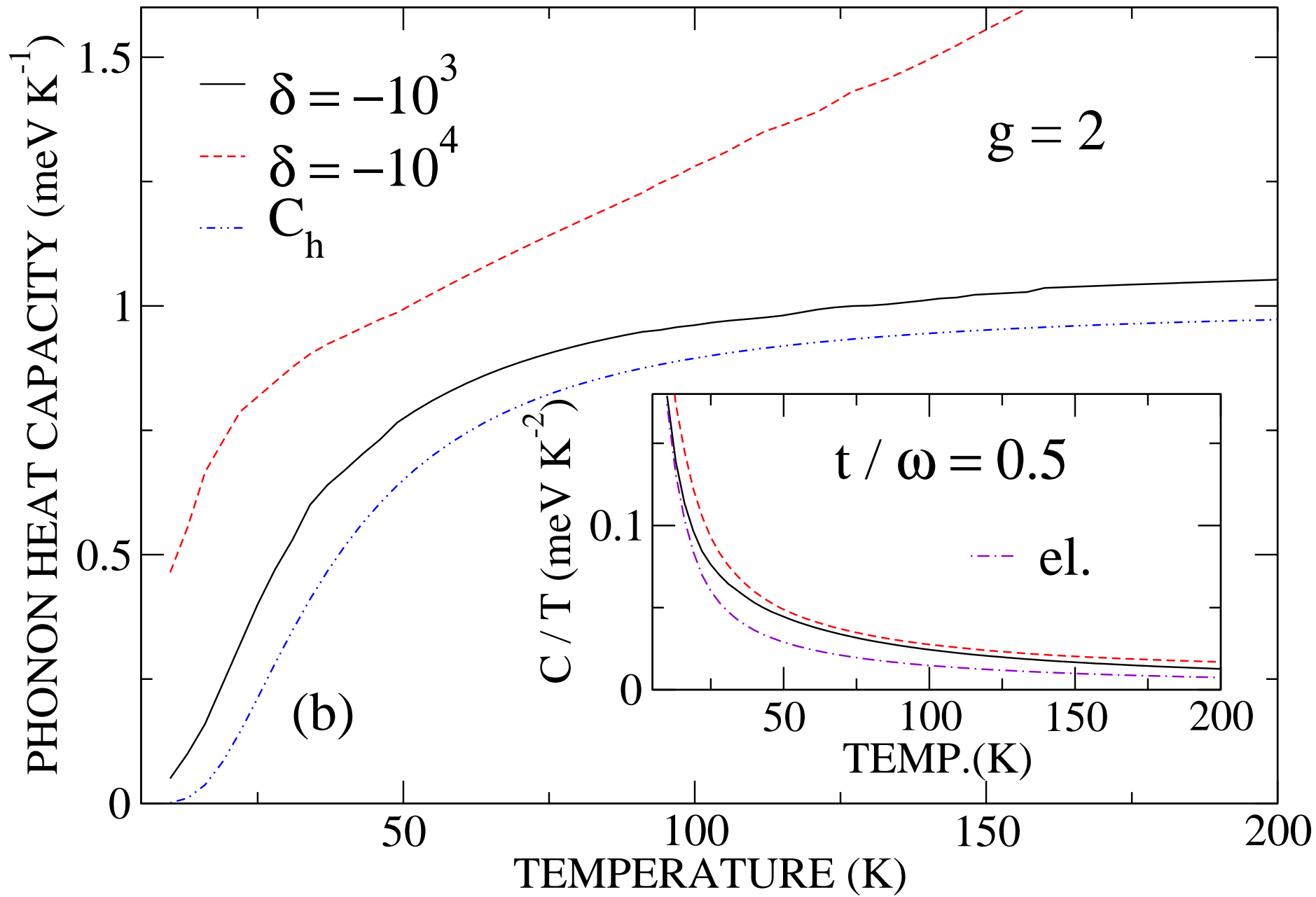
\end{document}